# Demodulation of Chaotic Signals Using Convolutional Neural Network

Mykola Kozlenko[1], Emrullah Demiral[2], Anton Yudhana[3]

[1]SoftServe Inc, mykola.kozlenko@ieee.org
[2]Department of Software Engineering, Karabuk University, Türkiye, emrullahdemiral@karabuk.edu.tr
[3]Department of Electrical Engineering, Universitas Ahmad Dahlan, Yogyakarta, Indonesia, eyudhana@ee.uad.ac.id

**Keywords:** Weak Signal Communications, Demodulation, Dynamic Chaos, Chaotic Signal, Machine Learning, Deep Learning.

**Abstract**

Chaotic modulation is an effective communication technique that exploits deterministic chaos to produce pseudo-random signals. A widely adopted approach involves modulation of the chaotic bifurcation parameter. This paper introduces a deep learning–based demodulation method for keying of the bifurcation parameter. It describes the architecture of the convolutional neural network and evaluates performance metrics for signals generated using the chaotic logistic map. The study assesses the bit error rate for binary signals and reports a bit error rate of 0.0819 for a bifurcation parameter deviation of 1.34% under additive white Gaussian noise at a signal-to-noise ratio of -13 dB (corresponding to a normalized signal-to-noise ratio of +20 dB). The results demonstrate the capability to detect chaotic patterns even when the specific patterns were not included in the training dataset.

## 1. Introduction

Deterministic chaotic signals exhibit complex, noise-like dynamics that arise from simple nonlinear deterministic rules (May, 1976). Despite being fully governed by mathematical equations, their evolution appears unpredictable due to high sensitivity to initial conditions. Such signals possess wideband and pseudo-random spectral characteristics, making them suitable as carriers in both analog and digital communication systems (Wang et al., 2017), (Tang et al., 2021). By exploiting these properties, chaotic modulation encodes information within the parameters or states of a chaotic system, offering an efficient wideband keying scheme that enhances transmission robustness and communication security (Quyen et al., 2011). It operates as a spread spectrum telecommunication system. Ergodic Chaotic Parameter Modulation (ECPM) is among the most widely used chaotic keying techniques (Leung et al., 2002). In this approach, digital information is represented by mapping data bits onto the bifurcation parameter (Borwein and Bailey, 2003, which directly controls the dynamic state of the transmitted chaotic signals (Abdullah and Ali, 2018). By varying this parameter, the system can produce distinct chaotic waveforms corresponding to different symbols, enabling efficient data transmission over a wide bandwidth. However, due to the sensitive dependence of chaotic systems on their parameters, ECPM requires sophisticated synchronization and demodulation algorithms, which significantly increase the complexity of the receiver design. The logistic map (Tsuchiya and Yamagishi, 1997) is a simple nonlinear rule capable of generating chaotic behavior. Applications based on the logistic map are widely employed across multiple domains, including information protection and digital communications (Mu and Liu, 2020), (de A. Kotaki and Luppe, 2020). The paper proposes a supervised machine learning-based approach for demodulating the bifurcation parameter of signals generated using the logistic map. The communication scenario is analyzed under additive white Gaussian noise (AWGN) conditions.

Replacing the logistic equation

$$\frac{dx}{dt} = rx(1-x) \tag{1}$$

with the quadratic recurrence equation

$$x_{n+1} = rx_n(1-x_n), \tag{2}$$

where $r$ is positive constant, parameter of the logistic map (Wolfram MathWorld, 2024).

In this study, chaotic signals are generated using the logistic map according to (2). Fig. 1 illustrates the bifurcation diagram for $r$ in the range from 2.8 to 4.0. Fig. 2 and 3 present the time-domain waveforms of the chaotic signals corresponding to $r$ = 3.70 and $r$ = 3.75, respectively.

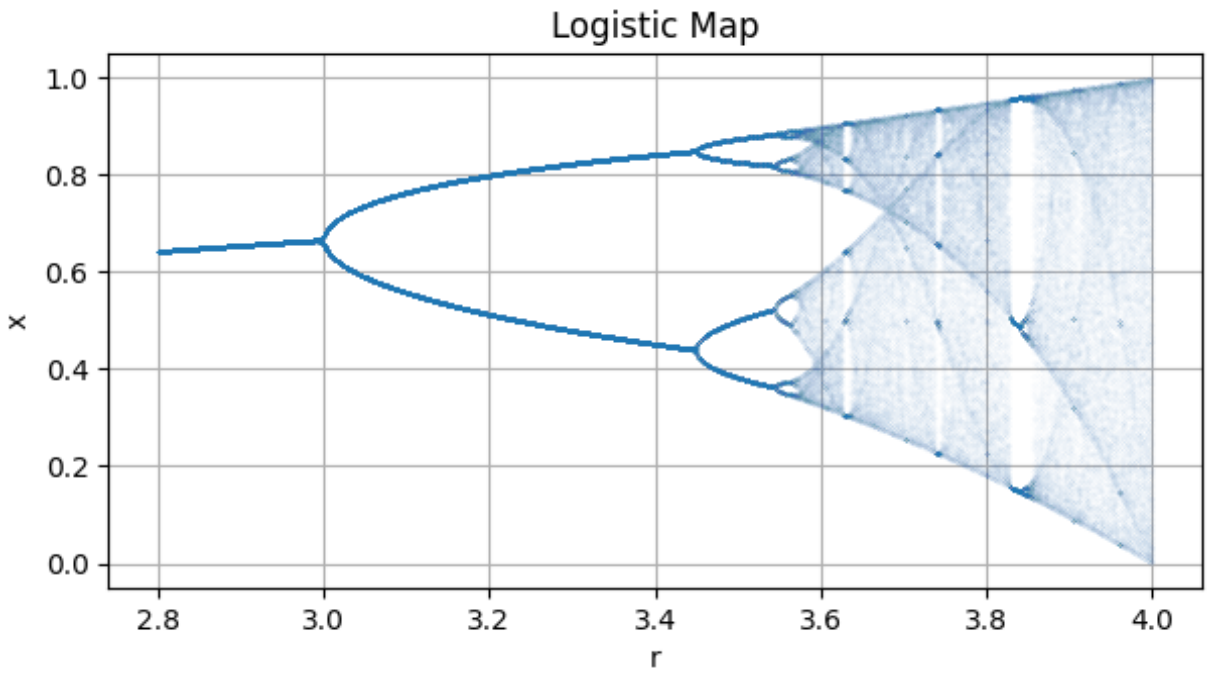

Figure 1. Bifurcation diagram of the logistic map.

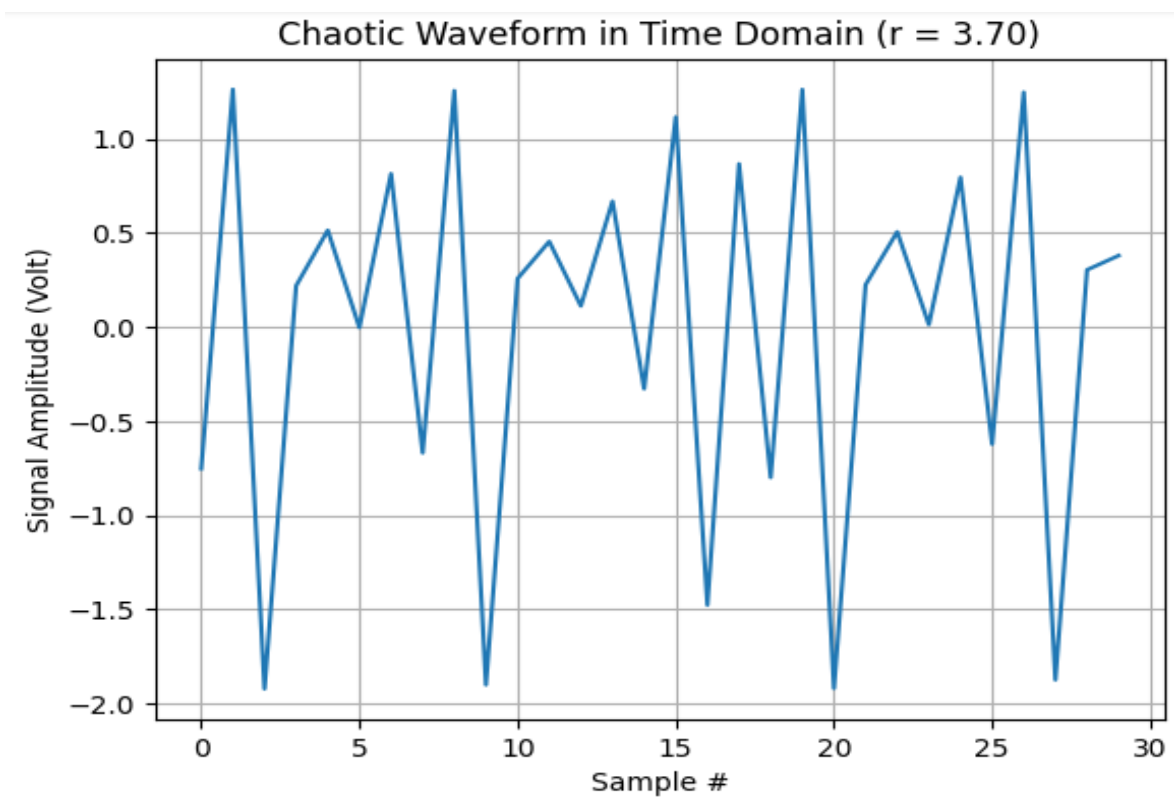

Figure 2. Time-domain waveform of the chaotic signal for $r$ = 3.70

This contribution has been peer-reviewed.
https://doi.org/10.5194/isprs-archives-XLVIII-4-W19-2025-79-2026

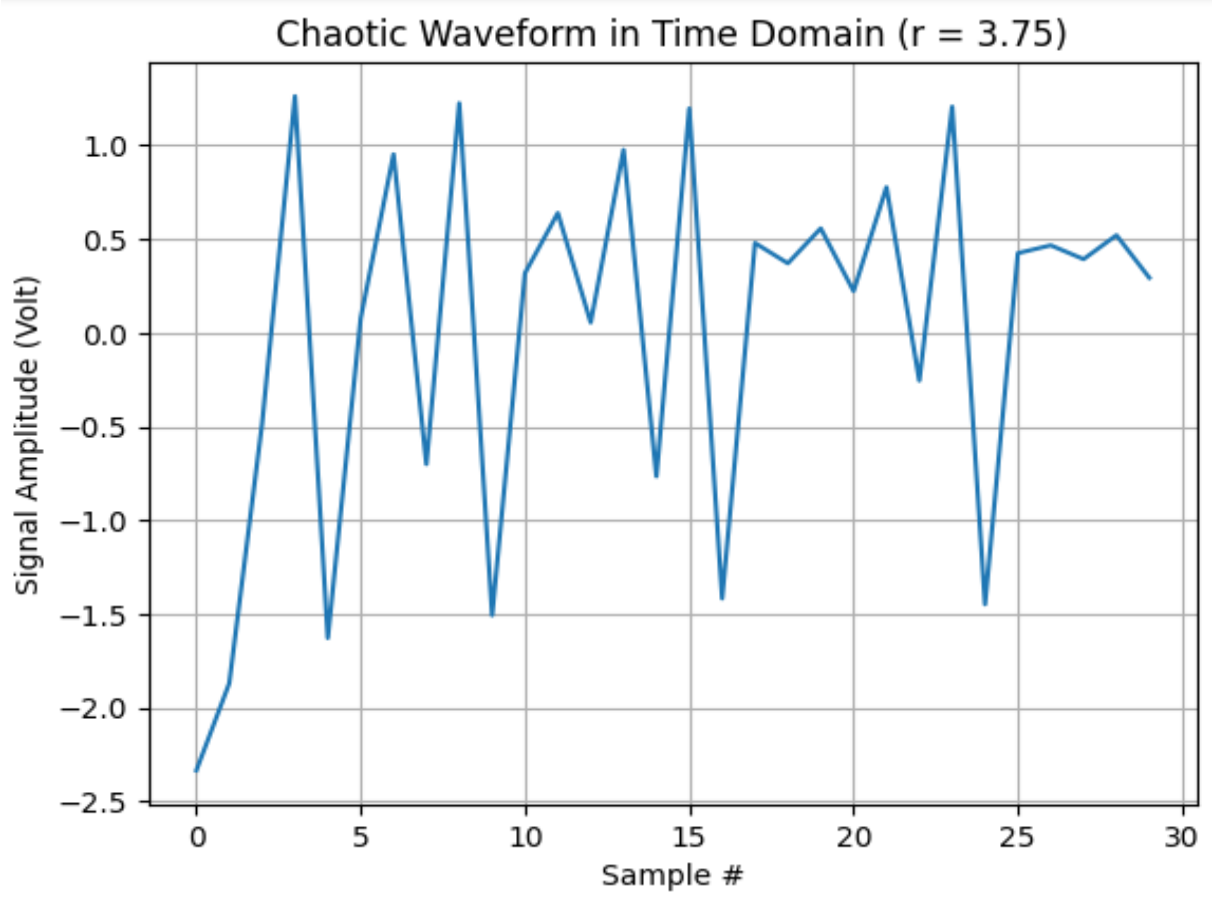

Figure 3. Time-domain waveform of the chaotic signal for $r = 3.75$

## 2. Related Work

Chaos has been applied in digital communications since the 90s (Hayes et al., 1993). The paper by Riaz and Ali (Riaz and Ali, 2008) provides an overview of chaotic communication systems and fundamental chaotic modulation schemes, including Chaos Shift Keying (CSK), Differential Chaos Shift Keying (DCSK), Multiplicative Chaos Modulation (MCM), and Additive Chaos Modulation (ACM). It also discusses synchronized chaotic systems and direct chaotic communication. The reference (Xing, 2020) introduces an information security inversion technique based on time series analysis. It covers chaotic signal demodulation methods, chaotic channel modeling, and analysis of chaotic signal interference channel problems. The paper (Xu et al., 2006) presents the application of a particle filter for demodulating the Chaotic Parameter Modulation (CPM) signal. In (Jin-feng and Jing-bo, 2007), the state equations of the unscented particle filter (UPF) for chaotic parameter demodulation were derived. Additionally, a novel modified algorithm based on the UPF for the demodulation process was proposed (Jin-feng and Jing-bo, 2007). The paper (Lazarovych et al., 2021) addresses the problem of demodulating weak radio signals for Phase Shift Keying (PSK) and Frequency Shift Keying (FSK) using Machine Learning (ML) techniques based on artificial neural networks. The paper (Li et al., 2018) formulates the demodulation of underwater communication signals as a classification problem and implements a machine learning-based demodulation scheme. The reference (Alvarez et al., 2004) demonstrates a security weakness in a communication method based on parameter modulation of a chaotic system and an adaptive observer-based synchronization scheme. It shows that the security can be compromised even without precise knowledge of the chaotic system properties (Alvarez et al., 2004). The paper (Ren et al., 2021) proposes the use of convolutional neural networks (CNN) within a deep learning framework to predict future symbols from the received signal, thereby reducing inter-symbol interference and achieving improved Bit Error Rate (BER) performance in chaotic baseband wireless communication systems. This paper extends the author's research published in (Kozlenko, 2022).

## 3. Dataset

Supervised machine learning approaches require a sufficiently large volume of training data. In many cases, obtaining an adequate amount of labeled data is a complex and resource-intensive task. In this study, we therefore use an artificially synthesized dataset generated with the same synthesis procedure as in our previous research. This approach allows us to maintain full control over data characteristics and class balance, ensuring reproducibility and consistency of the experiments. Although synthetic data may not capture all the nuances of real-world scenarios, it provides a valuable and flexible framework for model training and initial validation. The training dataset consists of 32000 records, the validation set contains 8000 records, and the test set comprises 10000 records.

In this study, the modulation scheme encodes digital data by varying the bifurcation parameter of a chaotic baseband carrier wave. The bifurcation parameter deviates by 0.05, ranging from 3.7 for a binary "0" to 3.75 for a binary "1" (corresponding to a percent deviation of 1.34%). The baseband bandwidth is 5512.5 Hz, with a sampling frequency of 11025 samples per second. Each bit has a duration of 0.3715 seconds and consists of 4096 samples, consistent with the JT65 protocol, resulting in a bit rate of 2.67 bit/sec. The signal-to-noise ratio (SNR) is -13 dB, the normalized SNR (Eb/N0) is +20 dB, and the spreading factor is +33 dB. The noise is modeled as additive white Gaussian noise (AWGN).

The example of the artificially synthesized signal waveform in the time domain at SNR value of +20 dB is shown in Fig. 4, its autocorrelation in Fig. 5, and the histogram in Fig. 6.

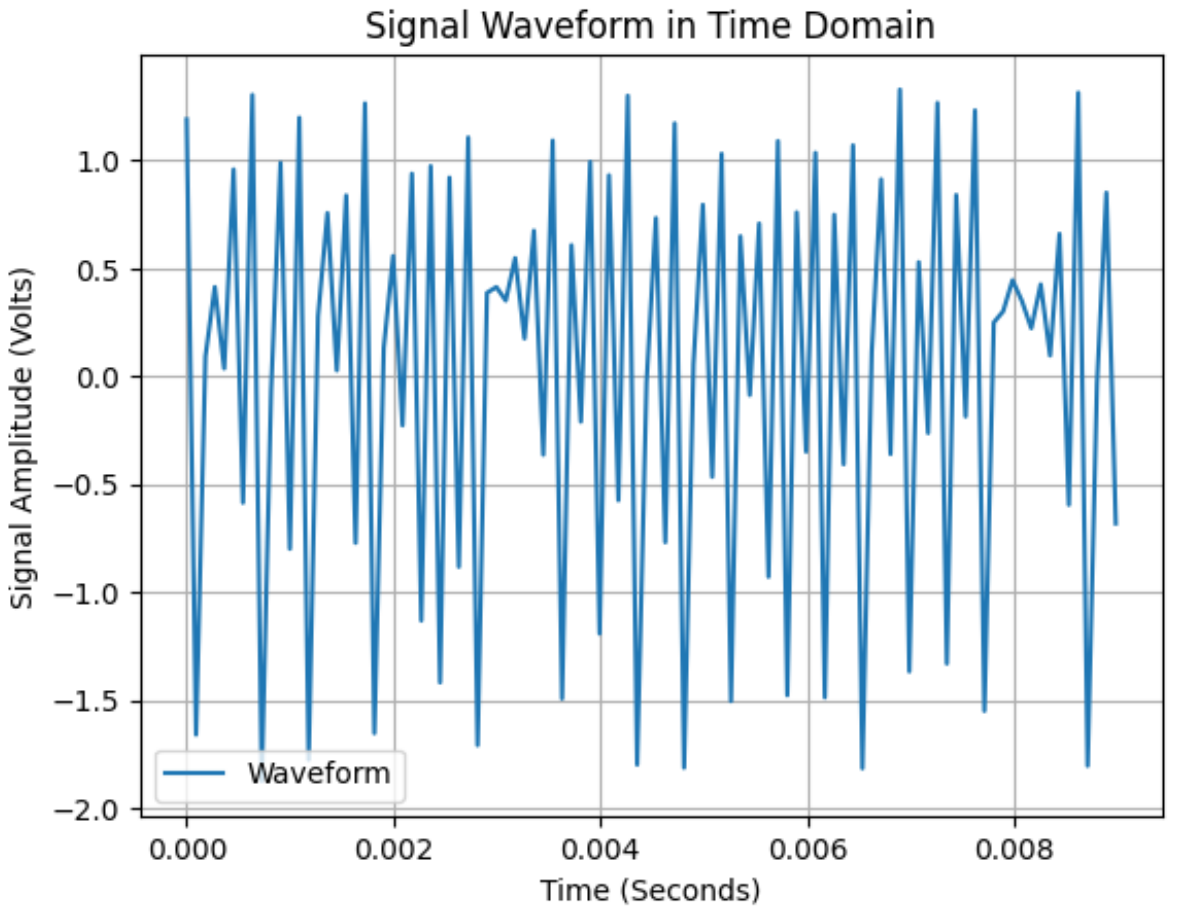

Figure 4. Time-domain signal waveform (selected 100 consecutive samples) at SNR = +20 dB.

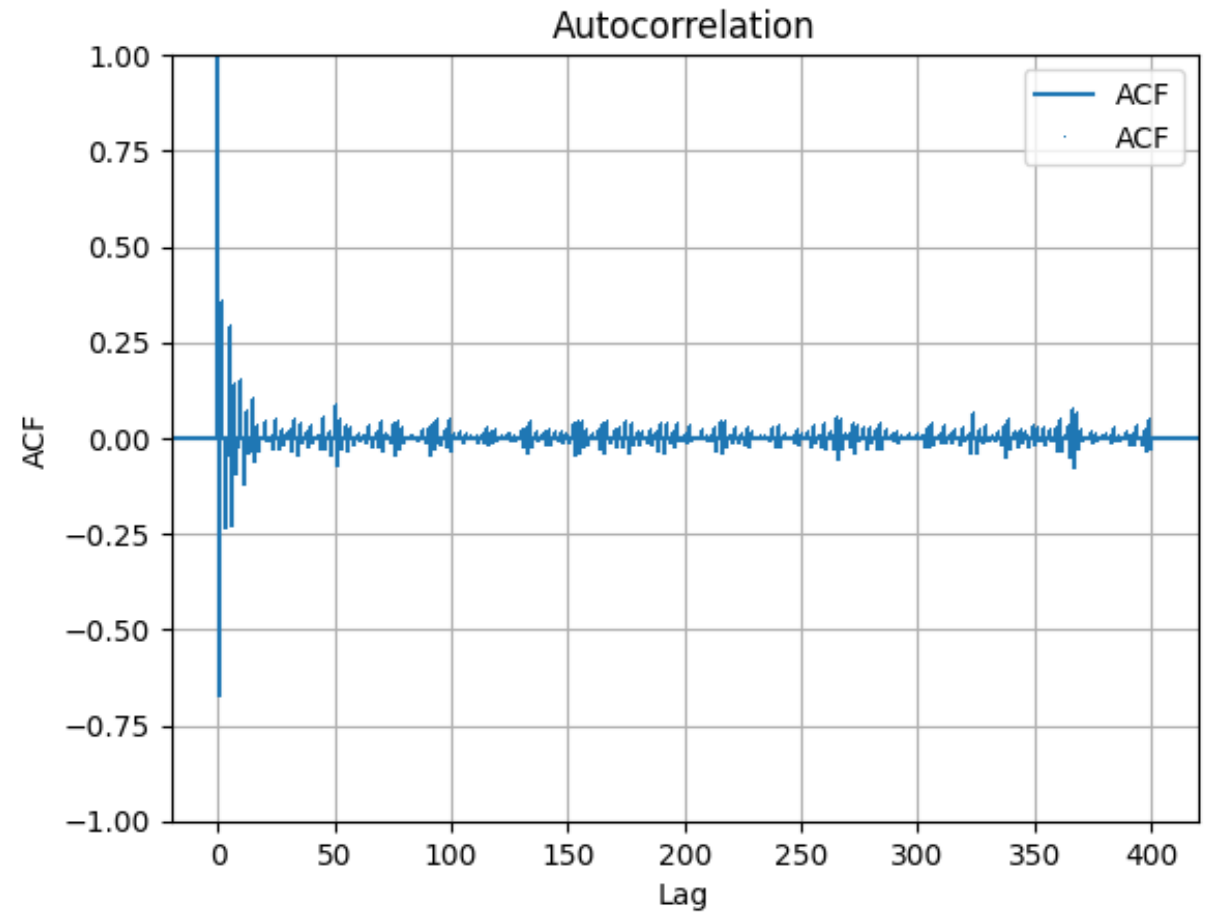

Figure 5. Autocorrelation of the signal (selected 400 consecutive samples) at SNR = +20 dB.

The Energy Spectral Density (ESD) of the synthesized signal was obtained using the Fast Fourier Transform (FFT) (Takahashi, 2019). The FFT was computed using the numpy.fft module. NumPy (van der Walt et al., 2011) is a vectorized Python library designed for scientific computing. The ESD for the SNR of +20 dB is shown in Fig. 7, the spectrogram in Fig. 8, and the scalogram based on wavelet (Guo, 2022) transform in Fig. 9. We used a Mexican hat ("mexh") wavelet (González-Nuevo, 2006).

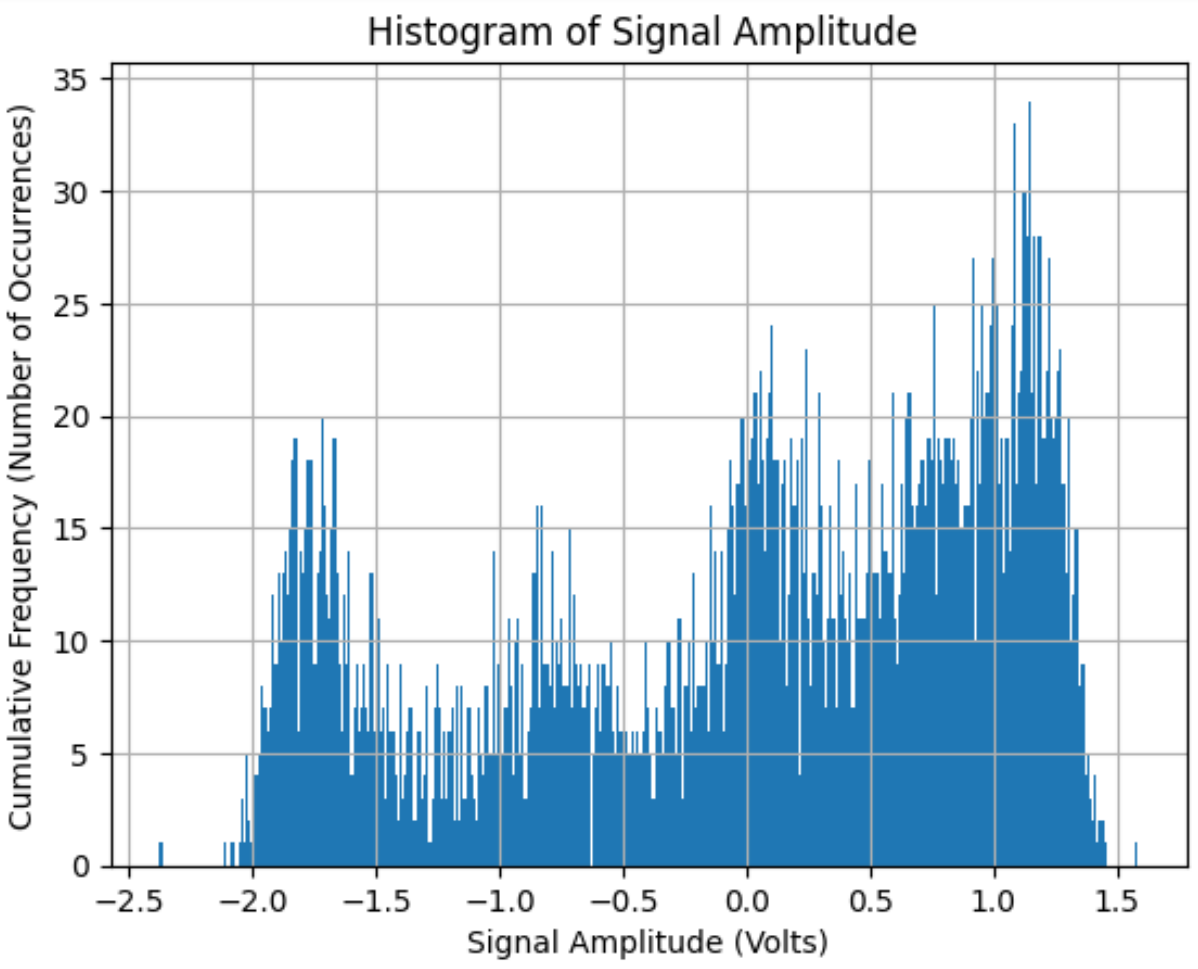

Figure 6. Histogram of the chaotic signal at SNR = +20 dB.

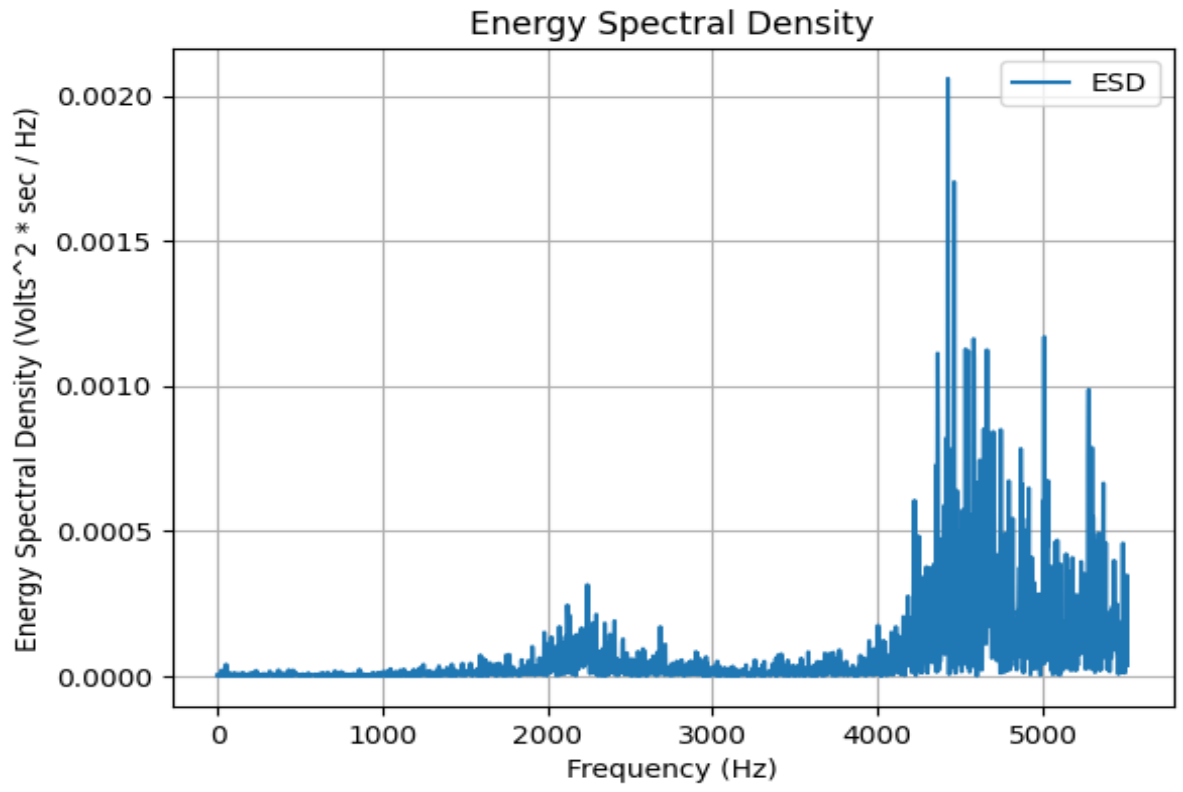

Figure 7. Energy Spectral Density of synthesized chaotic signal at SNR = +20 dB.

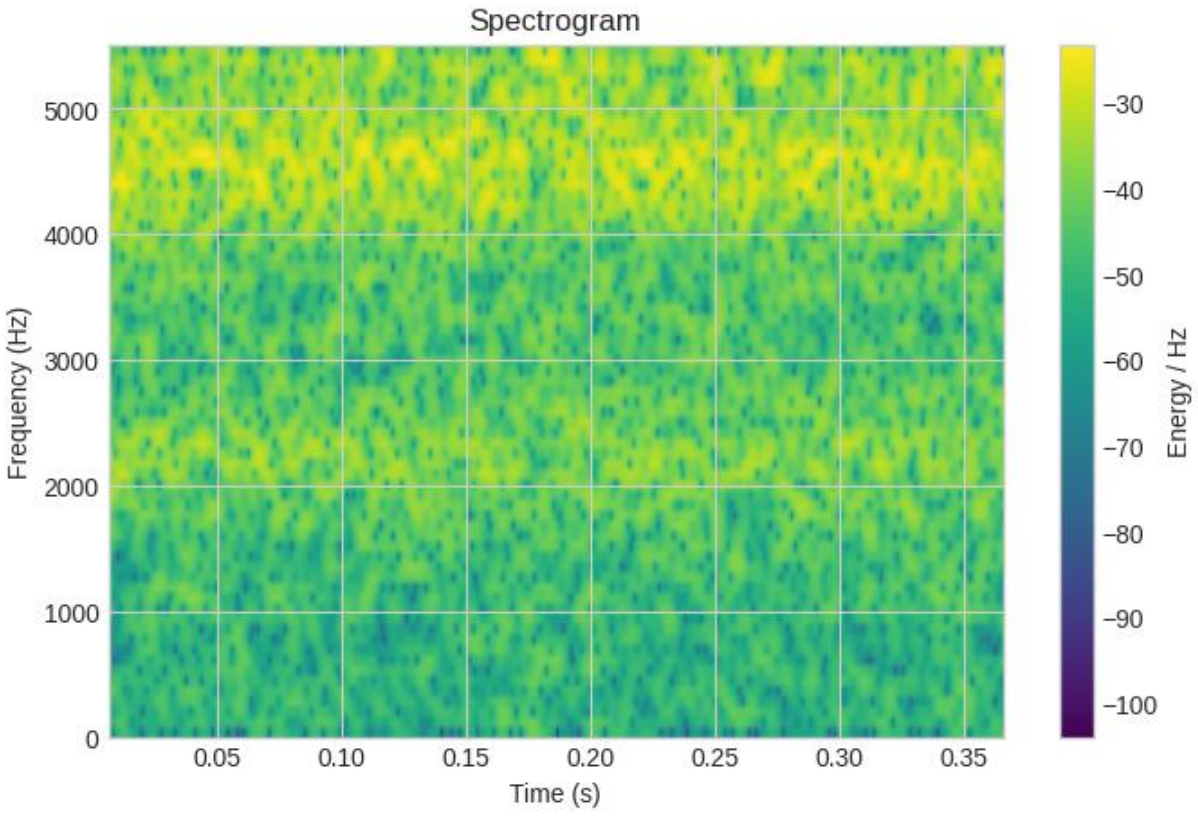

Figure 8. Spectrogram of the chaotic signal (r = 3.75, SNR = +20 dB).

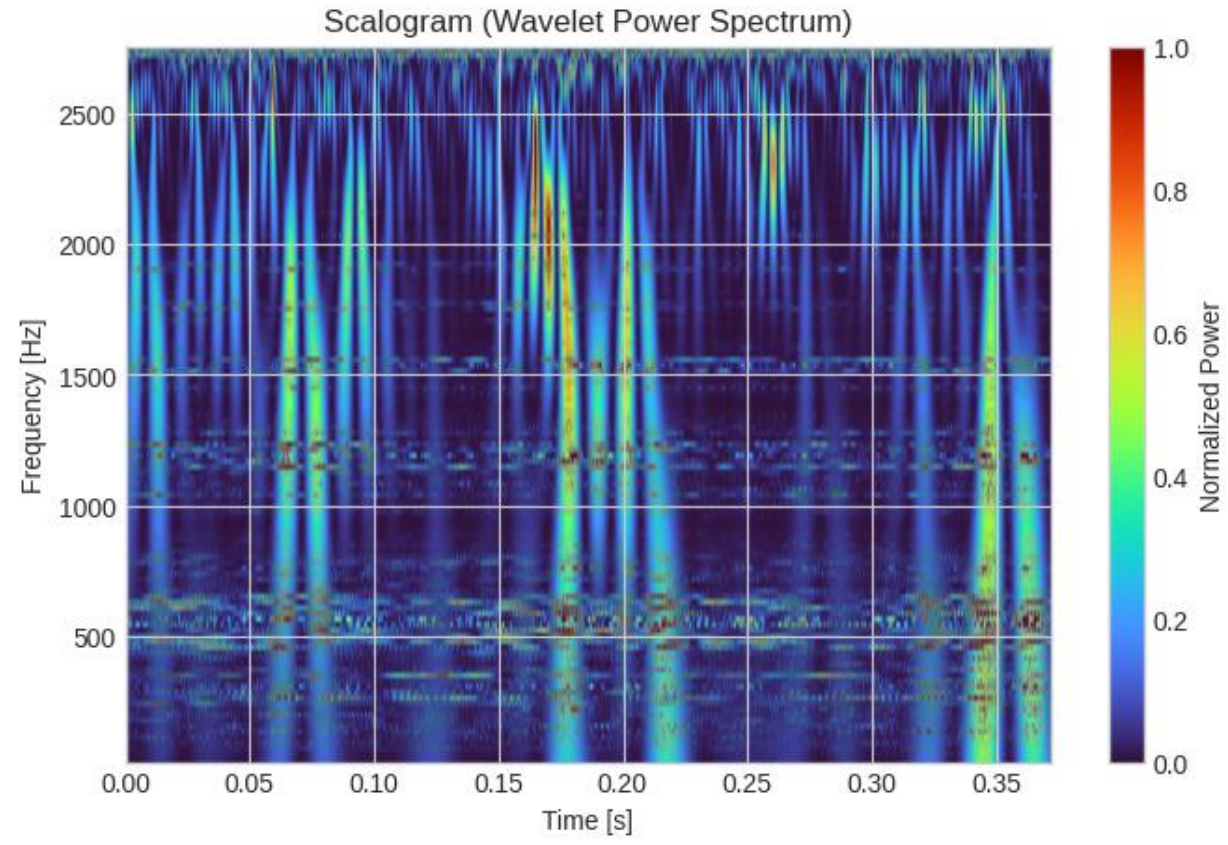

Figure 9. Scalogram of the chaotic signal (r = 3.75, SNR = +20 dB).

## 4. Methodology and Model Design

The neural network architecture (Fig. 10) consists of three one-dimensional convolutional layers, a hidden dense layer with ReLU activation, and an output layer with Softmax activation.

We utilize TensorFlow (Abadi et al., 2016) and Keras (Chollet et al., 2024) for model implementation. TensorBoard is employed to visualize training metrics and the neural network architecture. The model contains a total of 1,281,446 parameters, of which 1,248,164 are trainable and 33,282 non-trainable ones among them.

The architecture and hyperparameters are detailed as follows. Layer structure and data dimensions: input layer (4096), reshape layer (4096, 1), three convolutional blocks each consisting of batch normalization, one-dimensional convolution with filter size (16, 1), and max pooling with pool size (2, ). This is followed by a flatten layer (16,384), a dense layer (64) with ReLU activation, batch normalization (64), and an output dense layer (2) with Softmax activation. The optimizer used is Adam (built-in Keras implementation), the loss function is categorical cross-entropy, and evaluation metrics include class-wise error rate, overall accuracy, recall, and precision.

## 5. Training and Evaluation

Model training was conducted using the synthesized training dataset in Google Colaboratory (Colab) with an NVIDIA T4 Graphics Processing Unit (GPU) hardware accelerator based on the Turing Tensor Core technology with multi-precision computing. Training procedure takes approximately 36 seconds per epoch, 36 ms per step, 1.125 ms per one symbol interval. Each epoch contains 1,000 steps (batches). Each step contains 32 symbol intervals. The number of training epochs is 20. The training loss and accuracy as functions of the epoch number are presented in Fig. 11. The reported values correspond to those obtained at the end of each epoch.

We applied post-prediction evaluation to assess the model performance. The test dataset was processed through the prediction stage; in our case, it contains 10,000 symbol signals for each required signal-to-noise ratio. The predicted labels were then compared with the ground truth, and a confusion matrix was constructed. Based on the confusion matrix, the following class-wise and macro-/micro-averaged metrics were obtained: error rate, overall accuracy, true positive rate (recall, TPR), and positive predictive value (precision, PPV).

```
Layer (type)                 Output Shape              Param #
=================================================================
reshape (Reshape)            (None, 4096, 1)           0

batch_normalization (BatchNo (None, 4096, 1)           4

conv1d (Conv1D)              (None, 4096, 128)         2176

max_pooling1d (MaxPooling1D) (None, 2048, 128)         0

batch_normalization_1 (Batch (None, 2048, 128)         512

conv1d_1 (Conv1D)            (None, 2048, 64)          131136

max_pooling1d_1 (MaxPooling1 (None, 1024, 64)          0

batch_normalization_2 (Batch (None, 1024, 64)          256

conv1d_2 (Conv1D)            (None, 1024, 32)          32800

max_pooling1d_2 (MaxPooling1 (None, 512, 32)           0

flatten (Flatten)            (None, 16384)             0

batch_normalization_3 (Batch (None, 16384)             65536

dense (Dense)                (None, 64)                1048640

batch_normalization_4 (Batch (None, 64)                256

dense_1 (Dense)              (None, 2)                 130
=================================================================
Total params: 1,281,446
Trainable params: 1,248,164
Non-trainable params: 33,282
```

Figure 10. Architecture of the CNN model and the dimensionality of the data propagated through the network.

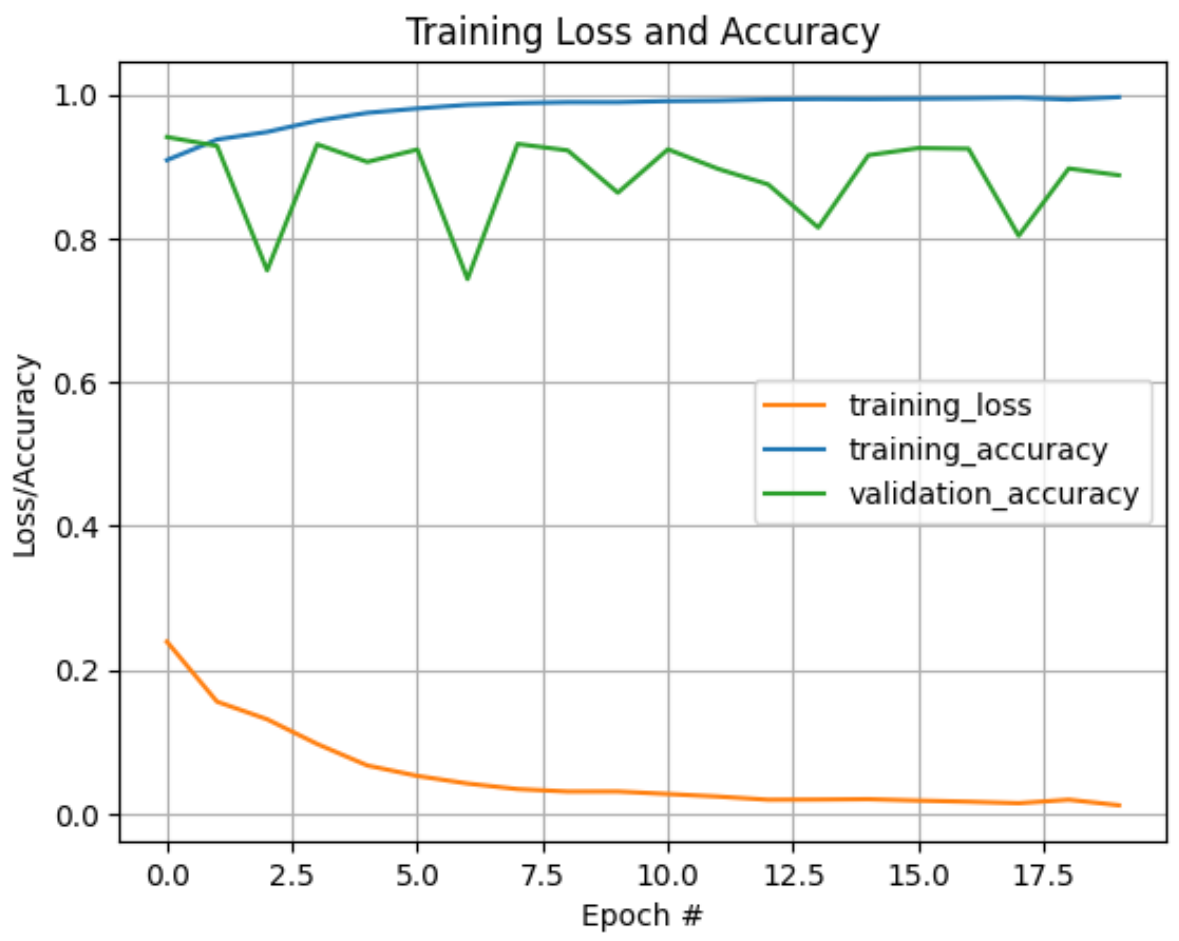

Figure 11. Training loss and accuracy versus epoch number.

## 6. Results

The detailed confusion matrix, which is a table used to evaluate the performance of a classification model by comparing predicted labels with true labels, is presented in Table 1. And the classification report (recall, precision, F1 score, support) for the test dataset is presented in Table 2, both at SNR value of -13 dB ($E_b/N_0$ = +20 dB).

| Predicted | True | | |
|---|---|---|---|
| | "0" | "1" | All |
| "0" | 4858 | 107 | 4965 |
| "1" | 1033 | 4002 | 5035 |
| All | 5891 | 4109 | 10000 |

Table 1. Confusion Matrix

| Class | Classification Metrics | | | |
|---|---|---|---|---|
| | Precision | Recall | F1-Score | Support |
| "0" | 0.82 | 0.98 | 0.89 | 4965 |
| "1" | 0.97 | 0.79 | 0.88 | 5035 |
| Accuracy | | | 0.89 | 10000 |
| Macro avg | 0.90 | 0.89 | 0.89 | 10000 |
| Weighted avg | 0.90 | 0.89 | 0.89 | 10000 |

Table 2. Classification Report

A commonly used metric (Sklar, 2001) for assessing the quality of any digital communication system is the dependence of the BER on the normalized signal-to-noise ratio ($E_b/N_0$). This relationship characterizes how efficiently a system can transmit information under noisy conditions and serves as a fundamental benchmark for comparing modulation and coding schemes. BER versus SNR and $E_b/N_0$ is presented in Table 3 (approximately +/- 1 dB around the average training SNR value) and Fig. 12 (+/- 3 dB around the average training SNR value). Results presented in Table 3 and Fig. 12 are from different runs of the pipeline and may differ a bit.

| SNR (dB) | −14 | −13 | −11.5 |
|---|---|---|---|
| $E_b/N_0$ (dB) | +19 | +20 | +21.5 |
| BER | 0.1154 | 0.0819 | 0.0393 |

Table 3. BER versus SNR and $E_b/N_0$

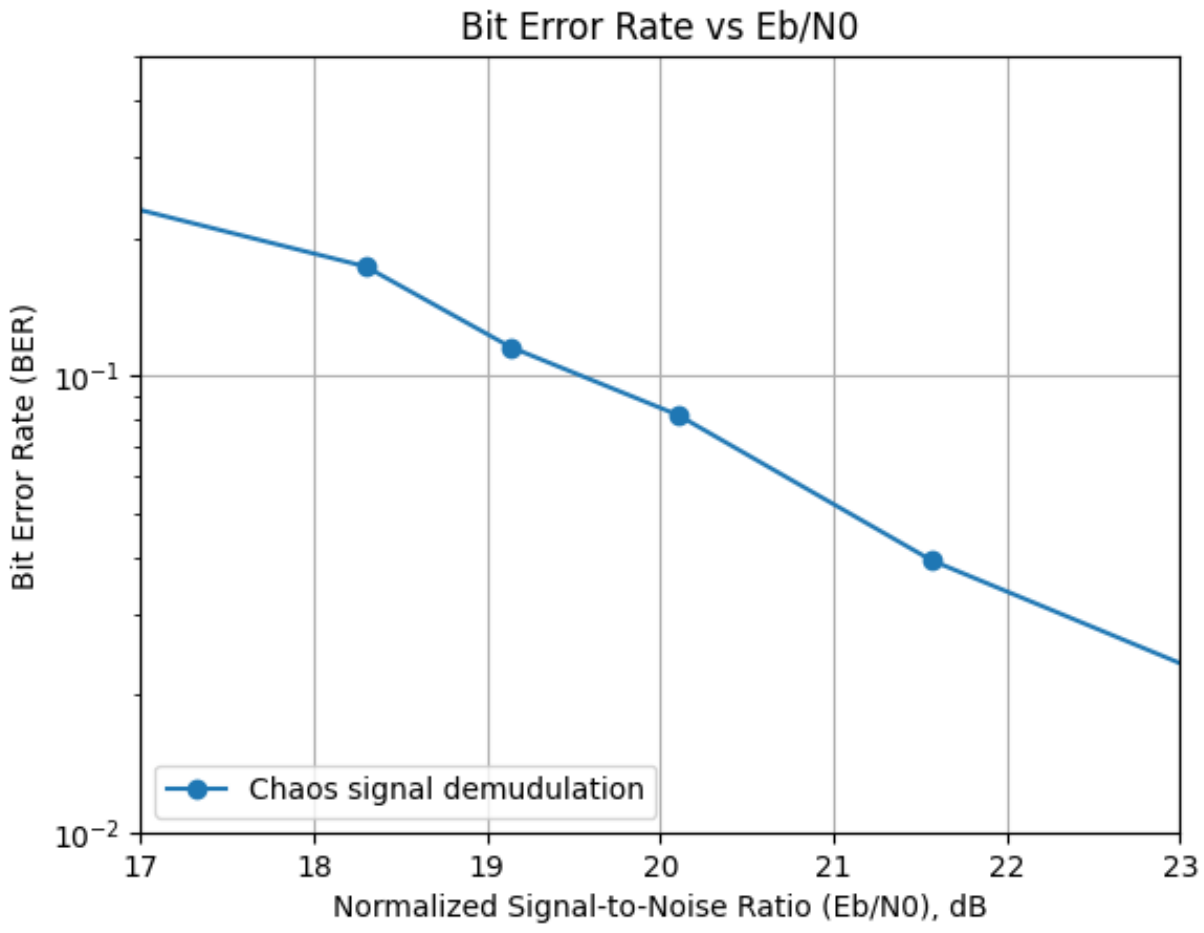

Figure 12. Bit error rate vs $E_b/N_0$.

Time complexity is one of the most critical aspects of real-time signal processing. We evaluated the run-time complexity of forward propagation by measuring the processing time of a single information symbol interval on the target hardware platform. The average processing time per information symbol on the NVIDIA T4 GPU platform is 156.25 μs (5 ms per batch of 32 information symbols). This value does not exceed the symbol interval duration. Therefore, real-time demodulation is feasible on the specified hardware platform.

## 7. Discussion

The objective of this study was to better understand the capability of ML algorithms to demodulate baseband chaotic signals. The findings support the hypothesis that chaotic signals generated using the logistic map can be effectively demodulated with an ML-based approach. The results further demonstrate that such demodulation remains feasible even in the presence of noise, which constitutes the main takeaway of this paper. This pattern of findings is consistent with prior studies on ML-based demodulation for FSK and PSK modulation schemes. The results provide a direct demonstration of chaotic signal demodulation within the ML framework. At least three limitations should be acknowledged. First, only signals generated by the logistic map were considered in this study. Second, a supervised machine learning approach was employed, requiring training on previously known bifurcation parameter values; an unsupervised approach may represent a more promising alternative. Third, the analysis was restricted to the AWGN case. Despite these limitations, the results suggest a practical implication: chaotic signals can be demodulated using a CNN-based approach. The Jupyter notebooks associated with this research are open-source and available for download from the GitHub repository (Chaos Research Team, 2024).

## 8. Suggestions for Further Research

For future research, it would be valuable to extend the present findings by investigating other types of chaotic maps. It would also be important to examine the potential of an unsupervised ML approach in the context of this task. In addition, the impact of other types of interference should be systematically studied. In terms of practical implementations, we are planning to extend our research (Kozlenko and Bosyi, 2018), (Kozlenko, 2015) data exchange in mobile robotics using noise-like signals, and in communications in wide-area monitoring systems and IoT (Kozlenko and Kuz, 2016).

## 9. Conclusion

The primary conclusion is that chaotic baseband signals generated using the logistic map within a bifurcation parameter modulation scheme can be successfully demodulated using a supervised machine learning approach. In summary, the paper reports the corresponding quality metrics, including a BER value of 0.0819 at an SNR of -13 dB (corresponding to a normalized SNR of +20 dB).

## Ethics Declaration

The authors have nothing to disclose.

## Generative AI Declarations

The authors have not employed any Generative AI tools.

## Acknowledgements

The author gratefully acknowledges the scientists of the Department of Information Technology of Vasyl Stefanyk Carpathian National University for their scientific guidance, valuable discussions, and technical assistance provided during the research.